\documentclass[twocolumn,prl, preprintnumbers,amsmath,amssymb,superscriptaddress,showpacs]{revtex4}
\usepackage{natbib} % First (see below)
\usepackage{graphicx}% Include figure files
\usepackage{dcolumn}% Align table columns on decimal point
\usepackage{bm}% bold math

\begin{document}

\title{Increasing the dimension in high-dimensional two-photon orbital angular momentum entanglement}
\author{J Romero}
\affiliation{School of Physics and Astronomy, SUPA, University of Glasgow, Glasgow G12 8QQ UK}
\affiliation{Department of Physics, SUPA, University of Strathclyde, Glasgow G4 ONG UK}
\author{D Giovannini}
\affiliation{School of Physics and Astronomy, SUPA, University of Glasgow, Glasgow G12 8QQ UK}
\author{S Franke-Arnold}
\affiliation{School of Physics and Astronomy, SUPA, University of Glasgow, Glasgow G12 8QQ UK}
\author{S M Barnett}
\affiliation{Department of Physics, SUPA, University of Strathclyde, Glasgow G4 ONG UK}
\author{M J Padgett}
\affiliation{School of Physics and Astronomy, SUPA, University of Glasgow, Glasgow G12 8QQ UK}

\begin{abstract} 
Any practical experiment utilising the innate D-dimensional entanglement of the orbital angular momentum (OAM) state space of photons is subject to the modal capacity of the detection system.  We show that given such a constraint, the number of measured, entangled OAM modes in photon pairs generated by spontaneous parametric down-conversion (SPDC) can be maximised by tuning the phase-matching conditions in the SPDC process.   We demonstrate a factor of 2 increase on the half-width of the OAM-correlation spectrum, from 10 to 20, the latter implying $\approx50$ -dimensional two-photon OAM entanglement.  Exploiting correlations in the conjugate variable, angular position, we measure concurrence values 0.96 and 0.90 for two phase-matching conditions, indicating bipartite, D-dimensional entanglement where D is tuneable.
\pacs{03.65.Ud, 03.67.Bg, 03.67.Mn}
\end{abstract}

\maketitle

Much attention has been directed to the two-dimensional state space of photon polarisation which provides both a conceptually and experimentally accessible playground\cite{Freedman1972, A.Aspect1981, Gisin2002RMP}. D-dimensional two-photon entanglement, wherein each photon is a D-level quDit taking on any of D possible values, is an even more fertile playground.   From a fundamental standpoint, higher-dimensional entanglement implies stronger violations of locality \cite{Collins2002, Dada2011} and is especially useful in the study of mutually unbiased bases in higher dimensions \cite{Wiesniak2011MUB}.  More relevant to practical applications, higher-dimensional entanglement provides higher information capacity \cite{Walborn2006QKD, Tittel2000QKD} and increased security and robustness \cite{Bechmann2000, Tittel2000QKD}.  Experimentally, D-levels in photons can be achieved by using the temporal and spectral degrees of freedom \cite{Silberhorn2009}, polarisation of more than one photon \cite{Bogdanov2004}, transverse spatial profile \cite{Walborn2006QKD}, position and linear momentum \cite{Howell2004EPR}, and angular position and orbital angular momentum \cite{LeachScience2010}.  

The entanglement of orbital angular momentum (OAM) in photons generated via spontaneous parametric down-conversion (SPDC) is firmly established theoretically and experimentally \cite{Franke-Arnold2002, A.Mair2001}.  The interest in OAM stems from its discrete and theoretically infinite-dimensional Hilbert space. Since the pioneering experiment of Zeilinger and co-workers ten years ago, OAM and it conjugate variable, angular position, has been steadily gaining ground as a mainstream variable in which to observe quantum correlations.   Bell-type and Leggett inequalities have both been violated in two-dimensional OAM subspaces analogous to the experiments done previously for polarisation \cite{J.Leach2009, Romero2010}.   The innate high-dimensional nature of OAM entanglement has been verified in an Einstein-Podolsky-Rosen (EPR) type experiment which measured both OAM and angular position \cite{LeachScience2010}.  A Bell-type inequality for higher dimensions has been recently violated using OAM states demonstrating experimental, two-photon, 11-dimensional entanglement \cite{Dada2011}.   The number of entangled OAM states that can be measured, i.e. the \emph{measurement spiral bandwidth} depends on both the detection capability and the number of OAM states that is generated by the down-conversion  process, i.e. the \emph{generation spiral bandwidth} \cite{Torres2003SB}.   The generation spiral spectrum  (i.e. the range of the D entangled OAM states and their respective weightings) of SPDC has been measured experimentally by a clever angular equivalent of the Hong-Ou-Mandel interferometer and bucket detectors \cite{vanExter2010SB}.  By exploiting the Fourier relationship between OAM and angle, the full generation spiral spectrum has been recovered and from this, the corresponding  generation spiral bandwidth has been deduced (albeit not projectively) and shown to strongly depend on the phase-matching conditions \cite{vanExter2010SB}.   

Any projective measurement of OAM, wherein the OAM of the signal and idler photons are directly measured using a mode-transformer (with a hologram or phase plate) and a single-mode fibre coupled to a photon-detector, is inherently sensitive to the radial field distribution \cite{VanExterFiltering2006, vanExter2010SB, Romero2011}.  Measuring the OAM spectrum in this manner will inevitably result to a spiral bandwidth that is different from the generation bandwidth \cite{vanExter2010SB}. However, for applications,  it is the measurement spiral bandwidth that represents the number of useable modes.  To increase this number, one can either optimise the detection system  or widen the OAM spectrum of the generated two-photon state.  These two are equally important, but optimising detection is fruitless if the OAM states are not being generated in the first place.  The detection geometry is more often fixed but can be designed optimally \cite{MiattoBoundsEPJ}.   The generation bandwidth can be modified by changing the characteristics of the pump beam \cite{Torres2003preparation, MiattoBoundsEPJ} , or by tuning the phase-matching conditions as shown previously  by temperature tuning a periodically poled potassium titanyl phosphate (PPKTP) crystal \cite{vanExter2010SB}.  More recently, pairs of OAM-entangled photons of arbitrary OAM value has been produced artificially by transferring path to OAM information. Although limited to a two-dimensional OAM subspace, the range of possible values are not constrained by OAM conservation \cite{Fickler2011}.

\begin{figure}[t!]
%\begin{center}
\includegraphics[width=8.8cm]{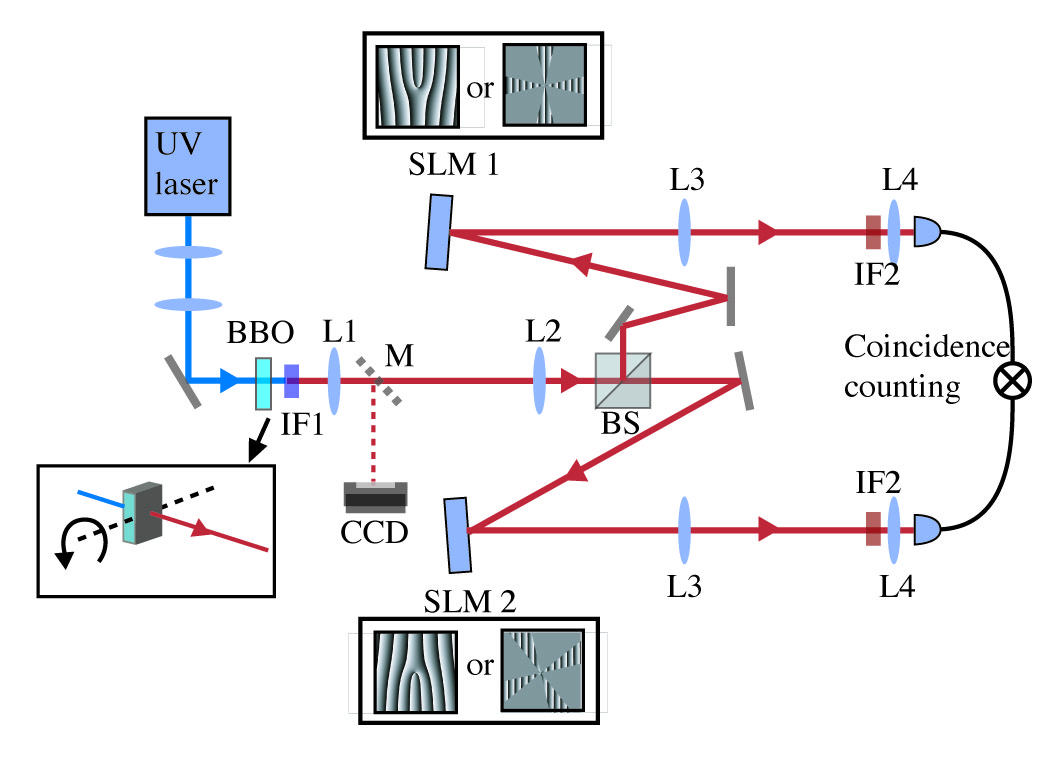}
\caption{\textbf{Experiment Scheme.}Photons from SPDC in a 5-mm long BBO crystal are measured in the OAM or angular position basis by programming either a forked diffraction hologram or an angular four-slit pattern on the SLMs (inset).  The SPDC phase-matching is changed by tilting BBO crystal about the axis shown (inset).  The CCD camera accessible via the flip-up mirror M allows us to derive a phase-matching parameter according to eq. \ref{PM}}
\label{setup}
%\end{center}
\end{figure}

\begin{figure}[t!]
%\begin{center}
\includegraphics[width=8.8cm]{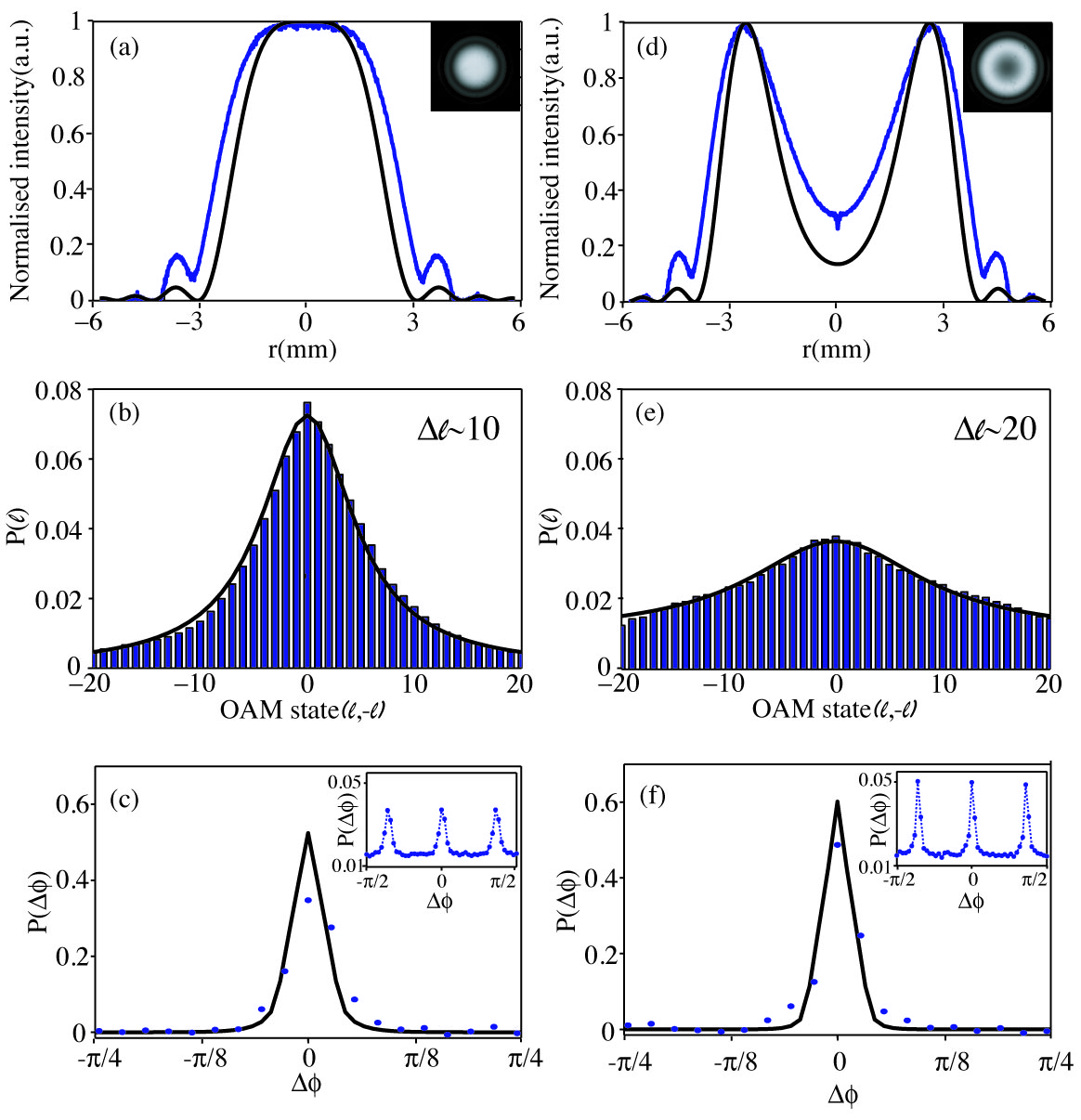}
\caption{\textbf{OAM and angular position measurements.} For collinear phase-matching, the far-field intensity profile (blue line) follows eq. \ref{PM} (solid black line) with $\alpha=0$ (a). The measured spiral spectrum  has a half-width $\Delta \ell\approx10$ (b). Angular position correlation as measured with four slits, when the signal and idler slits have a relative orientation of $\Delta\phi$  is shown in c inset. The central maximum (renormalised and background-subtracted) has a half-width of $\approx 12$ degrees (c).  For noncollinear phase-matching, with $\alpha=-2.2$ in eq. \ref{PM} (d),  the measurement spiral bandwidth is wider, with  $\Delta\ell\approx 20$ (e)  and the angular position correlation is narrower, with a half-width of $\approx 8$ degrees (f).   Blue dots and bars are experiment results, solid black lines are fits that demonstrate consistency with a Fourier relation between OAM and angle.}
\label{results}
%\end{center}
\end{figure}

Our entangled photons are generated from a 5-mm long beta-barium borate (BBO) crystal cut for collinear SPDC pumped by a collimated 355 nm pump laser of beam waist 500 $\mu$m (fig. \ref{setup}).  The pump beam is blocked by a longpass filter (IF1) after passing through the crystal. The crystal is mounted on a rotation stage which allows us to change the orientation of the crystal, consequently changing the phase matching from collinear to near-collinear. The signal and idler fields are incident on the same beam splitter and imaged by a telescope (lenses L1 and L2) to separate spatial light modulators (SLMs).  A flip-up mirror (M) is used to direct the light to a CCD camera positioned at the focal plane of L1 to allow us to capture the far-field intensity of the down-converted fields (figs. \ref{results} a,  d). The SLMs are imaged by lenses L3 and L4 onto the facets of single-mode fibres coupled to avalanche photodiodes(APD) for single photon detection.  Bandpass filters (IF2) of width 10 nm and centred at 710 nm placed in front of the fibres ensure that we measure signal and idler photons near degeneracy. The outputs of the APDs are connected to a coincidence circuit  and the coincidence rate is recorded as a function of the measurement states specified in the SLM.    

SPDC is the nonlinear interaction of three photons whose frequencies $\omega_j$ ($j$ stands for p-pump, s-signal or i-idler photon) are related as $\omega_{p}=\omega_{s}+\omega_{i}$. There is a range of wave vectors that will satisfy this energy conservation, and we can define an on-axis phase mismatch $\Delta k_z$ from the z-components of the wave vectors $\bold{k_j}$, $\Delta k_z=k_{p,z}-k_{s,z}-k_{i,z}$ \cite{Kleinman1968}.  The significance of phase-matching has been realised earlier on in the seminal paper of Kleinman \cite{Kleinman1968}, in which he calls SPDC as optical parametric noise. In harmonic generation, $2/\Delta k_z$ is the coherence length over which the three interacting fields remain in phase. In SPDC, $\Delta k_z$ has implications for efficiency (SPDC is brightest when $\Delta k_z\sim0$), but more importantly, determines the spectral distribution of the down-converted photons \cite{Boyd2003Book, Chen1988BBO}.  Theoretical treatment of phase-matching is complicated and several approximations have been made \cite{LawEberley2004, MiattoBoundsEPJ, Banaszek2009SPDC}, but it is easy to do in practice, either by tuning the temperature or angular orientation of the crystal \cite{Boyd2003Book}.  In the case of our bulk crystal, changing the angular orientation changes the index of refraction for the pump beam, and hence $\Delta k_z$ and
the far-field intensity profile of the down-converted fields.  The intensity profile $I$ we obtain mirrors the sinc-phase-matching term in SPDC and is fitted with the function,
 \begin{equation}
 I(r)={\rm sinc}^2\left(\frac{ar^2}{f^2} + \alpha\right)
 \label{PM}
 \end{equation}
 where $r$ is radial coordinate in the focal plane of a lens with focal length $f$ (L1), $\alpha=(|\bold{k}_p|-|\bold{k}_s| - |\bold{k}_i|)L/2$ is a phase-matching parameter which determines the opening angle of SPDC, and $a=(|\bold{k}_s| + |\bold{k}_i|)L/4n^2$, where $n$ is the refractive index for the signal and idler wavelengths and $L$ is the crystal length  \cite{PorsThesis}.  In the case where the transverse momentum of the photons is conserved, $\alpha$ is dominated by $\Delta k_z$, we take  this as a measure of our on-axis phase mismatch and $\alpha=0$ for the collinear case.
 
We measure both OAM and angle correlations for two different phase-matching conditions.  To measure OAM, we encode forked diffraction gratings of topological charge $\ell_s$ on one SLM and $\ell_i$ on the other. These holograms transform the incoming field to a fundamental mode which is the only mode that can be coupled to the fibres \cite{A.Mair2001}.  OAM is conserved in SPDC, hence we expect the OAM of the signal and idler photon to be anti-correlated, i.e. the coincidence count is high only  when $\ell_{s}=-\ell_{i}$.  Ideally, to measure correlations in angular position, we encode angular slits of width $\delta\phi$ centred at angle $\phi$ in both SLMs and rotate one with respect to the other, expecting high coincidence counts when the two slits are aligned  \cite{LeachScience2010}.  Because angle and OAM are Fourier-related \cite{Jha2008Fourier},  a wide spiral bandwidth means a correspondingly  narrow angular correlation which should be measured with a narrow angular slit.   This presents a limitation in practice because a narrow angular slit means less counts which are difficult to discern against the background. We solve this problem by using not one, but four narrow slits ($7$ degrees wide, almost twice as narrow as what was used previously \cite{LeachScience2010}) thereby enabling us to still measure tight angular correlations without sacrificing counts.   With one four-slit pattern oriented at $\phi_{s}$ and another oriented at $\phi_{i}$, we measure the coincidences as a function of $\Delta\phi=\phi_{s}-\phi_{i}$.  As a result of having four slits, our angular position coincidence curves has more than one maximum (figs. \ref{results} c , f insets) from which the width of the angular correlation can be derived. 

\begin{figure}[h!]
%\begin{center}
\includegraphics[width=8.8cm]{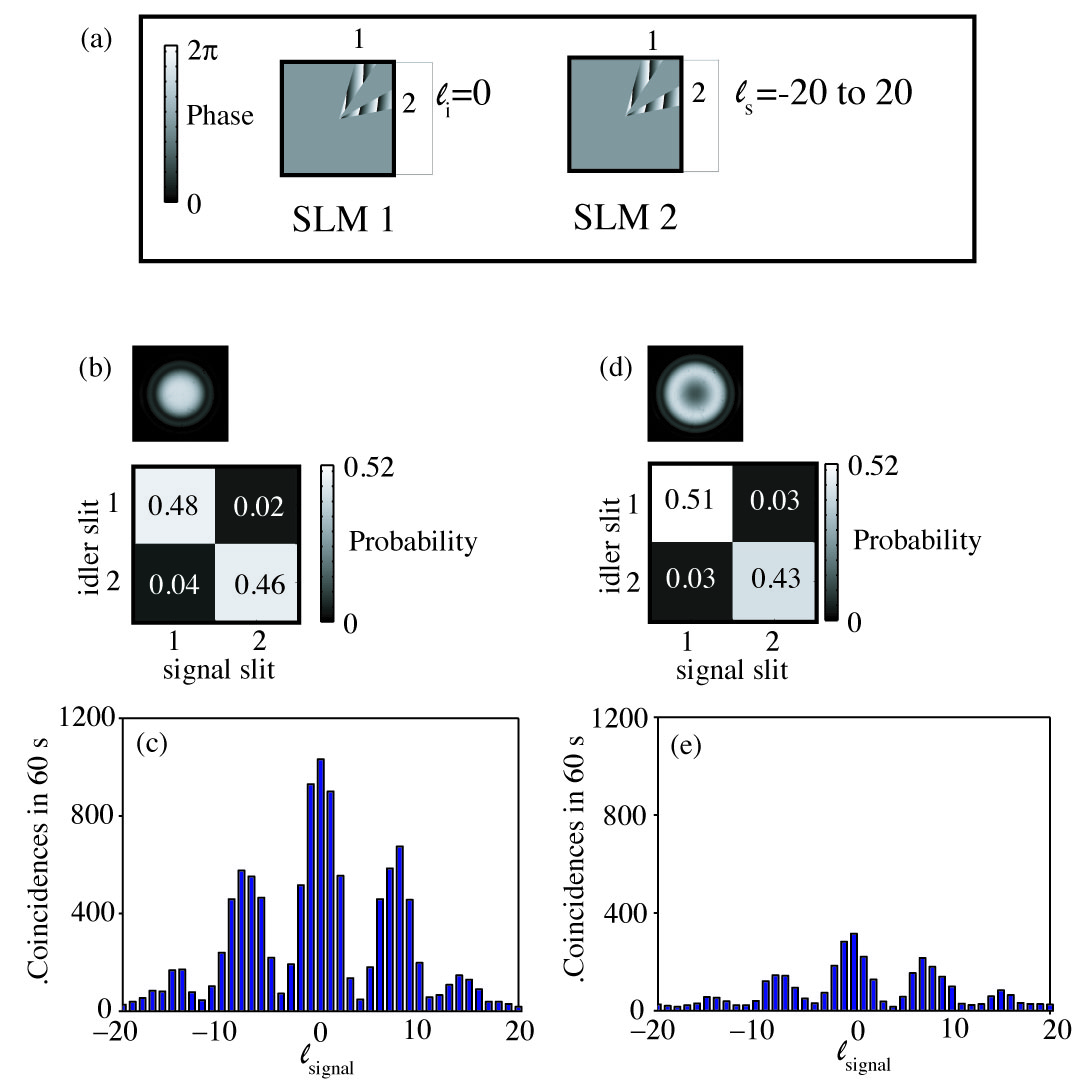}
\caption{\textbf{Concurrence measurements.}  We encode two-slit patterns (width $18^o$, separated by $45^o$) in SLM1 and SLM2 with the corresponding OAM values shown (a).  With only one slit in each SLM (1 or 2), we verified the strong angular position correlation. For $\alpha=0$, we only get high coincidences when both SLMs have slits with the same angular position (b).  The  measured concurrence is 0.96 (c).  We show similar angular position correlation for $\alpha=-2.2$, (d) and  measure a concurrence of 0.90 (e). }
\label{concurrence}
%\end{center}
\end{figure}

OAM and angular position measurements for two different phase-matching conditions are shown in fig. \ref{results}.  We define the measurement spiral bandwidth, $\Delta\ell$, as the full width at half-maximum of the measured spiral spectrum.  For collinear phase-matching, $\alpha=0$ in eq.\ref {PM} (fig. \ref{results} a),  we find $\Delta\ell\approx10$ (fig. \ref{results} b) and the corresponding half-width of the  central peak in the angular position coincidence curve is $12^o$ (fig. \ref{results} c).   With the addition of an on-axis phase mismatch, $\alpha=-2.2$ in eq.\ref {PM}, such that the opening angle of the spot is slightly larger ($\approx1.1^o$ from propagation axis to the first minimum, compared to $0.9^o$ for $\alpha=0$ ) and there is a central dip in the intensity distribution  (fig. \ref{results} d), we find $\Delta\ell\approx20$ (fig. \ref{results} e).  The half-width of the corresponding angular position correlation is narrower, as expected from the Fourier relationship \cite{Jha2008Fourier,Jack2011Uncertainty}, and is $8^o$.   The solid black lines in fig. \ref{results} b and e are Lorentzian and are empirical fits to our data. Using these fits, we were able to calculate the expected angular correlation from the Fourier relation, with the added consideration that our angular masks has a finite slit width (solid line in fig. \ref{results} e and f ).

Analytical treatment of the effect of phase-matching on the generation spiral bandwidth is difficult especially for the noncollinear case.  Instead we offer a simple geometrical argument that will elucidate why the spiral spectrum widens as we tune the phase-matching. This involves the concept of the optical \'{e}tendue, $E=A\Omega$, where $A$ is the near-field beam area and $\Omega$  is the solid angle subtended by the beam in the far-field \cite{Markvart2008Etendue, VanExterFiltering2006}.  In the treatment of noise in laser amplifiers, the \'{e}tendue normalised with respect to the wavelength $\lambda$, $E/\lambda^2$, is the number of transverse modes that can be supported, $E$ acts as a measure of the quantum states in a beam \cite{YarivBook, Markvart2008Etendue}.  $E$ is more often invoked in the discussion of light collection, but is equally applicable in the case of SPDC where light is instead, being emitted. Regardless of the phase-matching, $A$ is the same in our experiment: the SLMs are in the near-field of a particular plane in the crystal and minute changes to crystal orientation (typically 1/20 of a degree) do not change the image on the SLMs.  However, this changes the far-field opening angle, $\Omega$ increases for $\alpha=-2.2$, hence the number of transverse (both azimuthal and radial)  modes emitted increases.  In setting up an SPDC experiment, this has important practical implications,  one should ensure that the detection \'{e}tendue is greater than the generation \'{e}tendue to maximise the overlap between the pump and detection modes. Using the Klyshko picture as a guide, where the detected signal ( or idler) mode is back-propagated to the crystal, reflected off the crystal, and propagated to the other detector \cite{Klyshko1988}, the overlap is  maximised by keeping all corresponding far-field solid angles and near-field beam diameters in the signal and idler arms the same.   The reason we use two lenses to image the crystal onto the SLMs, instead of one, is to match the far-field angles in both arms, and this gives us an effective detection system.

Measuring the entanglement of D dimensions is not as straightforward as measuring the entanglement of 2-dimensional systems. We can violate a Bell inequality for higher dimensions as implemented in \cite{Dada2011}, but this is difficult for ${\rm D}\sim 40$ because the intensity mask reduces the count rates considerably.   Instead, we exploit the Fourier relationship, or complementarity between OAM and angle \cite{Jha2008Fourier,Barnett1990quantum}.  When a photon passes through an angular aperture we can observe interference in the OAM distribution of the signal (or idler) field, the modulation of which depends on the spiral spectrum of the photons \cite{Jha2010}.  We can encode angular two-slit patterns on the SLMs (slits 1 and 2. fig. \ref{concurrence} a) and measure the resulting OAM interference when SLM1 (idler) is set to measure $\ell_{i}=0$ and the value of $\ell_{s}$ on SLM2 is scanned from $-\ell_{max}$ to $\ell_{max}$ ($\ell_{max}=20$ in our case).    It has been shown that the visibility of the resulting interference pattern is the same as the concurrence (ranges from 0 to 1, 1 being the maximally entangled case) of the two-qubit density matrix written in the angular position basis \cite{Wootters1998, Jha2010}.    We verify strong angular position correlation in  fig. \ref{concurrence} b and d where we have measured the coincidences when we encode only one slit (of width $18$ degrees) on each SLM for both phase-matching conditions. As expected, we only get appreciable coincidences when we encode the same slit positions for both SLMs.  Ideally the diagonals should be 0.5, but due to imperfect alignment we get small probabilities off the diagonal. The interference of the two-slit patterns in fig. \ref{concurrence} a with their corresponding OAM values leads to a modulation in the coincidences which can be measured in the OAM basis.  Fig. \ref{concurrence} c and e show the coincidences for $\alpha=0$(b inset) and $\alpha=-2.2$(d inset).   The measured concurrence is $0.96$  for $\alpha=0$ and $0.90$ for $\alpha=-2.2$, demonstrating that we indeed have entangled angular qubit states for both phase-matching conditions. We emphasise that although we have resorted to quantifying the entanglement in the angular position basis, the measurements made in fig. \ref{concurrence} b and d (strong angular position correlation) and fig. \ref{concurrence} c and e (interference in the OAM basis) can be produced simultaneously only by OAM-angular position entangled sources.  

In conclusion, we have demonstrated that minute changes to the angular orientation of a bulk BBO crystal ($\approx$1/20 of a degree) widens the OAM measurement spiral spectrum and narrows the angular position correlation, as a consequence of phase-matching in SPDC.   We have designed our detection system guided by the concept of the optical \'{e}tendue, and we have achieved a measurement bandwidth (FWHM) of 20, which translates to $\approx 50$ useable OAM modes \cite{MiattoBoundsEPJ}.  The fact that we can generate \emph{and} detect this number of modes, implying experimental 50-dimensional two-photon OAM entanglement,  points to the possibility of new experiments such as loophole-free Bell test experiments \cite{Brunner2010closing}, superdense coding \cite{Pati2005Superdense} and quantum walks \cite{Schreiber2010} where a higher-dimensional space is desirable.

We acknowlede the UK EPSRC, DARPA, Wolfson Foundation, the Royal Society, and EU HIDEAS for funding,  Hamamatsu for SLM loan and M. van Exter for useful discussions. 
\bibliographystyle{apsrev}

\end{document}